\documentclass[aps,prl,twocolumn,showpacs,superscriptaddress,groupedaddress]{revtex4}  % for review and submission
\usepackage{graphicx}  % needed for figures
\usepackage{dcolumn}   % needed for some tables
\usepackage{bm}        % for math
\usepackage{amssymb}   % for math

\usepackage{color}

\newcommand{\planck}{{\it Planck} }
\newcommand{\mnu}{\sum m_\nu}
\newcommand{\neff}{{\rm N}_{\rm eff}}
\newcommand{\msterile}{m_{\nu,{\rm sterile}}^{\rm eff}}

\begin{document}

\title{No new cosmological concordance with massive sterile neutrinos}
%\title{Too early for a new concordance cosmology involving non standard neutrino properties}

\author{Boris Leistedt}
  \email{boris.leistedt.11@ucl.ac.uk}
  \affiliation{Department of Physics and Astronomy, University College London, London WC1E 6BT, U.K.}

\author{Hiranya V. Peiris}
  \email{h.peiris@ucl.ac.uk}
  \affiliation{Department of Physics and Astronomy, University College London, London WC1E 6BT, U.K.}

\author{Licia Verde}
  \email{liciaverde@icc.ub.edu}
  \affiliation{ICREA (Instituci\'o Catalana de Recerca i Estudis Avan\c{c}at) \& ICC, Institut de Ciencies del Cosmos, Universitat de Barcelona (UB-IEEC), Marti i Franques 1, Barcelona 08028, Spain} 
  \affiliation{Institute of Theoretical Astrophysics, University of Oslo, 0315 Oslo, Norway. \\}

\date{\today}

\begin{abstract}
It has been claimed recently that massive sterile neutrinos could bring about a new concordance between observations of the cosmic microwave background (CMB), the large-scale structure (LSS) of the Universe, and local measurements of the Hubble constant, $H_0$. We demonstrate that this apparent concordance results from combining datasets which are in significant tension, even within this extended model, possibly indicating remaining systematic biases in the measurements. We further show that this tension remains when the cosmological model is further extended to include significant tensor modes, as suggested by the recent BICEP2 results. Using the Bayesian evidence, we show that the minimal $\Lambda$CDM model is strongly favoured over its neutrino extensions by various combinations of datasets. Robust data combinations yield stringent limits of $\sum m_\nu\lesssim0.3$~eV and $m_{\nu,{\rm sterile}}^{\rm eff} \lesssim 0.3$~eV at $95\%$ CL for the sum of active and sterile neutrinos, respectively. 
\end{abstract}

\maketitle

The temperature fluctuations of the cosmic microwave background (CMB), as measured by the \planck satellite \citep{Planck_cosmo_2013}, have yielded sub-percent level constraints on the cosmological parameters of the vanilla $\Lambda$CDM model. However, the primary CMB temperature fluctuations only indirectly probe the growth of cosmic structure, and it is therefore essential to complement it with observations large-scale structure (LSS) such as galaxy clusters, weak lensing, and clustering measurements. The first cosmological results from the \planck satellite have revealed a $\sim 2$$\sigma$ tension between CMB temperature measurements and the Sunyaev-Zel'dovich (SZ) cluster abundances \citep{Planck_SZ_2013}, mainly in terms of $\sigma_8$, the linear-theory mass dispersion on a scale of $8h^{-1}$ Mpc. A similar tension is observed with the X-ray cluster counts 
\citep{Vikhlinin_etal_2009}.

Massive neutrinos can potentially alleviate this tension because they suppress power in the clustering of matter at late times. They are an appealing solution since solar and atmospheric experiments have already provided evidence for their mass, with room for extra sterile species, supported by anomalies in short baseline and reactor neutrino experiments (for reviews of particle physics constraints, see {\it e.g.}, Refs.~\cite{PhysRevD.86.010001, GonzalezGarcia:2012sz, Conrad:2013mka, Abazajian:2012ys}). Cluster abundances, galaxy surveys and weak lensing are sensitive to the total neutrino mass, either from active neutrinos $\mnu$ (the total mass from active species), or sterile neutrinos $\msterile$ (a effective parameter which  connects to actual neutrino masses in the context of specific models, see {\it e.g.}, Ref.~\cite{Lesgourgues_Pastor_2013}). In addition, an extra parameter $\neff$ can be introduced to denote the effective number of relativistic species, in which case $\neff>3.046$ (the standard number) is referred to as ``dark radiation'', and is also appealing as it could alleviate the tension between \planck and local $H_0$ measurements \cite{2013PDU.....2..166V}. 

A number of recent studies have carried out joint analyses of various data combinations to conclude that these tensions are resolved within a new concordance model which implies non-standard neutrino parameters \cite{Battye_Moss_2013, Wyman_etal_2013, Hamann_Hasenkamp_2013, Beutler_etal_2014, Giusarma:2014zza}.  Ref.~\cite{Battye_Moss_2013} argued that combining the CMB with lensing or SZ cluster measurements reveals evidence for non-zero neutrino mass in both the active and sterile neutrino scenarios. Refs.~\cite{Wyman_etal_2013, Hamann_Hasenkamp_2013} claimed that sterile neutrinos could reconcile  \planck  with LSS data, in particular with the X-ray cluster abundances \citep{Vikhlinin_etal_2009} and the latest constraints on $H_0$  \citep{Riess_etal_2011}. By combining the CMB with shear and redshift space distortion (RSD) measurements, Ref.~\cite{Beutler_etal_2014} found hints of non-zero masses for active neutrinos. Finally, Refs.~\cite{Zhang:2014dxk, Dvorkin:2014lea} further claimed that sterile neutrinos could resolve a potential tension between \planck and BICEP \cite{bicep2_2014} constraints on $r_{0.002}$, the tensor-to-scalar ratio at $k=0.002$ Mpc$^{-1}$.  

Although these conclusions are not universally accepted \cite{Planck_cosmo_2013, Feeney:2013wp, Hu:2014qma, Verde:2013cqa, Efstathiou:2013via}, tension between the datasets may indeed point to new physics. Alternatively, tension may also indicate remaining systematic biases in the measurements, which can have substantial impact on cosmological parameter measurements at the level of precision achieved by current data. Consequently, new physics in the neutrino sector is only a viable solution if the extra parameters eliminate the tension between datasets seen in the standard concordance cosmology, and is robustly confirmed by a variety of datasets. In this {\it Letter}, we show that sterile neutrinos do not relieve the tension between \planck\ and X-ray and SZ clusters, or with local measurements of $H_0$.  Further, we show that the extended neutrino models are not preferred over the minimal model by any data combination, and that robust combinations of current measurements prefer low neutrino masses $\mnu$,  $\msterile \lesssim 0.3$~eV. 

\begin{figure*}
\hspace*{-6mm}\includegraphics[width=18cm]{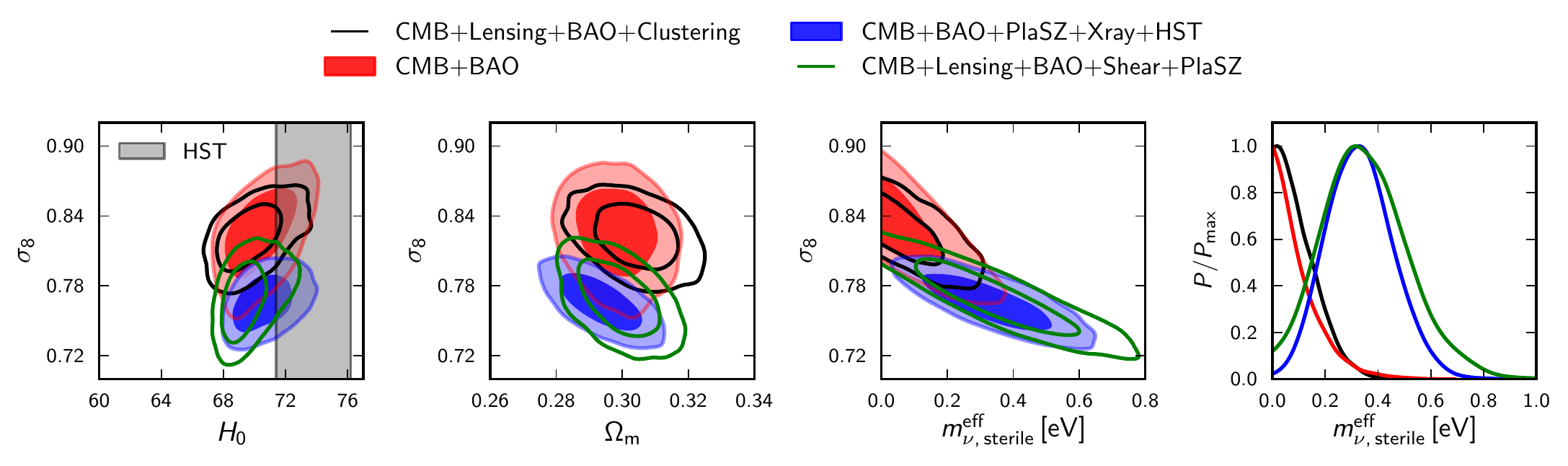}
\caption{Constraints on the $\Lambda$CDM+${\rm N}_{\rm eff}+\msterile$ model, showing that non-zero sterile neutrino mass is only favoured as a result of a tension between the CMB and cluster data (PlaSZ, X-ray) in the $\sigma_8$--$\Omega_m$ plane, and the degeneracy between $\sigma_8$ and neutrino mass.}
\label{fig:noconcordance}
\end{figure*}

\textbf{Data and methods.} We use CosmoMC \cite{Lewis:2002ah} to constrain the parameters of the $\Lambda$CDM model extended with active (+${\rm N}_{\rm eff},\mnu$) and sterile (+${\rm N}_{\rm eff},\msterile$) neutrinos, using combinations of the following datasets. 
 \underline{CMB}: the \planck CMB temperature likelihood \cite{Planck:2013kta}, combined with Wilkinson Microwave Anisotropy Probe (WMAP) polarisation \cite{Bennett:2012fp}, and high-$\ell$ temperature spectra from Atacama Cosmology Telescope (ACT) and South Pole Telescope (SPT) \citep{Keisler_etal_2011, Das_etal_2013, Reichardt_etal_2012}.
 \underline{Lensing}: the CMB lensing likelihood from \planck \cite{Ade:2013mta}.
 \underline{BAO}: the Baryon Acoustic Oscillations (BAO) measurements from 6dF \cite{Beutler_etal_2011}, Sloan Digital Sky Survey (SDSS) DR7 \cite{Padmanabhan:2012hf}, WiggleZ \cite{Blake:2011en}, and Baryon Oscillation Spectroscopic Survey (BOSS) DR11 \cite{Anderson_etal_2013}. 
\underline{Shear}: the weak lensing tomographic analysis from Canada-France Hawaii Telescope Lensing Survey (CFHTLenS) \citep{Kilbinger:2012qz}.
\underline{PlaSZ}: the \planck SZ cluster abundances \citep{Planck_SZ_2013}.
\underline{RSD}: the RSD measurements from BOSS \citep{Beutler_etal_2013, Beutler_etal_2014}.
\underline{Xray}: X-ray cluster mass function constraints \citep{Vikhlinin_etal_2009}.
\underline{HST:} the $H_0$ measurement using supernovae by the {\it Hubble Space Telescope } \citep{Riess_etal_2011}. 
\underline{Clustering}: the 3D galaxy power spectrum from WiggleZ \cite{Blake:2010xz, Parkinson:2012vd}, and the power spectrum of the reconstructed halo density field derived from Luminous Red Galaxies (LRG) in SDSS DR7 \cite{Reid_etal_2010}, both up to $k=0.2$ $h$Mpc$^{-1}$. Note that we only use either the power spectrum or the BAO measurement from each dataset. 

Finally, we use the Evidence ratio (or Bayes factor), which gives the relative odds of two models correctly describing the observations, under the assumption of equal {\it a priori} model probabilities (see {\it e.g.}, Refs.~\cite{Cox, Verde:2013cqa} and references therein). We calculate $\ln [E_{\Lambda{\rm CDM}}/E_{\rm ext.}]$, the logarithm of the Evidence ratio of the $\Lambda$CDM model divided by that of the the extended neutrino models; thus, positive numbers favour the minimal model. In practice, since the models are nested, we compute Evidence ratios with the Savage-Dickey Density Ratio, and we use Kernel Density Estimation (KDE) to process the MCMC chains and reliably compute the marginalised posterior distributions at the $\Lambda$CDM values ($\mnu=0.06~\textrm{eV}, \msterile=0.0~\textrm{eV}, \neff=3.046$). The errors are calculated by jackknifing the KDE parameters. For all parameters, we consider the same prior ranges as the official {\it Planck} analysis \citep{Planck_cosmo_2013}. However, the Bayes factors only depend on the neutrino parameters since we consider nested models. Specifically, we assume uniform priors in $[0,5]$, $[0,3]$ and $[3.046, 10]$ for $\mnu$, $\msterile$ and $\neff$, respectively, and we impose $\msterile/(\neff-3.046)<7$~eV to avoid a degeneracy between very massive neutrinos and cold dark matter.

\textbf{No new concordance with sterile neutrinos.}   Fig.~\ref{fig:noconcordance} shows constraints on the $\sigma_8$--$\msterile$ plane for several data combinations, including those used by Refs.~\cite{Battye_Moss_2013, Wyman_etal_2013, Hamann_Hasenkamp_2013}. Our minimal dataset is CMB+BAO, since adding BAO to CMB does not shift the contours but constrains the matter density $\Omega_m$ and reduces the error-bars (as expected for consistent datasets). However, the addition of the PlaSZ or X-ray clusters, which prefer lower $\sigma_8$, shifts the contours significantly (by more than 2$\sigma$) outside the region allowed by CMB+BAO. This clearly indicates that the addition of sterile neutrinos to the $\Lambda$CDM model does not bring the CMB and cluster measurements into agreement. Note that the active scenario (not shown here) leads to similar results and tension, and does not yield concordance within the extended model either. Thus we may conclude that the tension must be resolved either by considering systematics in one or more of the relevant datasets, or else by new physics other than the introduction of massive (active or sterile) neutrinos.  This is confirmed by the Bayes factor, presented in the first section of Table~\ref{tab:bayesfactors_tension}, showing that the extended models are not preferred over the minimal $\Lambda$CDM model even in the presence of a tension.

\begin{table}\def\arraystretch{1.1}
	\centering 
	\caption{Evidence ratios $\ln [E_{\Lambda{\rm CDM}}/E_{\rm ext.}]$ between the minimal $\Lambda$CDM model and the extended neutrino models, in the active and sterile scenarios, showing that the extended models are not favoured by any data combination. In particular, the upper part refers to the ``tension'' data combinations of Fig.~\ref{fig:noconcordance}, whereas the lower part corresponds to more robust data combinations (details in text), for which marginalised constraints are presented in Tables~\ref{tab:active} and \ref{tab:sterile}.}
	\begin{tabular}{lccc}
		&	Active	&	Sterile 	 	\\\hline
CMB+BAO+PlaSZ+Xray+HST  & $1.52  _{-0.33}   ^{+0.16} $   & $-0.16  _{-0.35}   ^{+0.39} $ \\ 
CMB+Lensing+BAO+Shear+PlaSZ  & $3.77  _{-0.09}   ^{+0.10} $   & $1.05  _{-0.55}   ^{+0.26} $ \\ 
\hline
CMB+BAO  & $4.42  _{-0.05}   ^{+0.04} $   & $3.10  _{-0.14}   ^{+0.07} $ \\ 
CMB+Lensing+BAO  & $4.64  _{-0.09}   ^{+0.03} $   & $2.99  _{-0.05}   ^{+0.06} $ \\ 
CMB+Lensing+BAO+Clustering  & $4.70  _{-0.00}   ^{+0.02} $   & $3.35  _{-0.13}   ^{+0.09} $ \\ 
CMB+Lensing+BAO+Clusters  & $4.65  _{-0.19}   ^{+0.10} $   & $2.61  _{-0.23}   ^{+0.21} $ \\ 
CMB+Lensing+BAO+Shear  & $4.32  _{-0.16}   ^{+0.10} $   & $2.10  _{-0.41}   ^{+0.21} $ \\ 
CMB+Lensing+BAO+RSD  & $4.14  _{-0.19}   ^{+0.10} $   & $1.81  _{-0.09}   ^{+0.11} $ \\ \hline
	\end{tabular}
	\label{tab:bayesfactors_tension}
\end{table}

Cluster cosmology is currently limited by modelling rather than statistical uncertainties \cite{Planck_SZ_2013}; thus, error-bars on the X-ray, SZ and optical clusters data used in Fig.~\ref{fig:noconcordance} and in Refs.~\cite{Battye_Moss_2013, Wyman_etal_2013, Hamann_Hasenkamp_2013, Beutler_etal_2014, Zhang:2014dxk, Giusarma:2014zza, Dvorkin:2014lea} may need to be significantly increased to account for additional potential systematics. The calibration of the mass-observable relation is critical for deriving robust cosmological constraints from clusters, and is complicated by uncertainties in mass measurements and  the selection functions (see {\it e.g.}, Refs.~\citep{Vikhlinin_etal_2009, Rozo_etal_2013}). Constraints on $\sigma_8$ from PlaSZ clusters are sensitive to assumptions and uncertainties in the modelling, as investigated in Ref.~\cite{Planck_SZ_2013}, and there are indications of a systematic mismatch between masses obtained via weak lensing compared with SZ masses \citep{VonDerLinden2014}. The error bars on $\sigma_8 (\Omega_m)^\beta$ from X-ray clusters used in Ref.~\cite{Battye_Moss_2013} should be enlarged to account for confirmed sources of systematic uncertainties \cite{Vikhlinin_etal_2009}.  Interestingly, it was shown that the mass calibration by Ref.~\cite{Rozo_etal_2013b} from a self-consistent analysis of X-ray, SZ, and optical scaling relations is consistent with a minimal flat $\Lambda$CDM model with no massive neutrinos (1.7$\sigma$), and is a better fit to additional data ({\it e.g.} $H_0$).  
Finally, the model dependence of these cluster constraints in the context of non-standard models has not been investigated; therefore it is unclear whether they can be used in a joint analysis in the context of such extended models. 

If, after further investigation of such systematic effects, PlaSZ and X-ray clusters remain in tension with CMB+BAO, this tension cannot be simply resolved by adding sterile neutrinos.

\textbf{Constraints on neutrino masses from robust datasets.} 
We now investigate the constraints obtained on neutrino masses when combining datasets which are compatible and have been demonstrated to be robust to modelling uncertainties. Recent works using galaxy power spectra have obtained tight constraints on the mass of active neutrinos ({\it e.g.}, Refs.~\cite{Thomas:2009ae, Giusarma_etal_2013, Riemer-Sorensen:2013jsa}), and also showed that it could help in breaking degeneracies with the freedom in the primordial power spectrum from inflation \cite{deputter_etal_2014}. For \underline{Clustering} data, we use the power spectra from SDSS DR7 (reconstructed halo power spectrum) and WiggleZ (galaxy power spectrum), truncated at $k=0.2$ $h$Mpc$^{-1}$ in order to avoid non-linear scales, marginalising over the galaxy bias. For \underline{Shear} data, we use the tomographic weak gravitational lensing analysis by the CFHTLenS \citep{Kilbinger:2012qz}, which were shown to be usable in neutrino extensions of $\Lambda$CDM \cite{Beutler_etal_2014}. For the \underline{Clusters} data, we use the thermal SZ measurements from cross-correlation of the CMB with X-ray clusters \citep{Hajian_etal_2013}, which are the most recent cluster-derived cosmological constraints. They rely on cross-correlations, and were also demonstrated to be robust to the choices in the modelling and data (tested with \planck and WMAP). We jointly use the \planck \underline{CMB} temperature and \underline{Lensing} power spectra (to probe the growth of structure with the CMB) with the \underline{BAO} constraints (to constrain $\Omega_m$). Finally, we also use the \underline{RSD} measurements from BOSS \citep{Beutler_etal_2013}.

\begin{table}
	\centering 
	\caption{Marginalised $95\%$ CL constraints on the $\Lambda$CDM+${\rm N}_{\rm eff}+\mnu$ model from a variety of robust LSS datasets with the \planck CMB temperature and lensing measurements. These datasets  are not in tension and tightly constrain the mass of active neutrinos.}
	\begin{tabular}{lccc}
		&	$\mnu$[eV] 	&	$\neff$ 	 	\\\hline
	 CMB+BAO	&	$<$0.23		&	$<$3.88	 	\\
	 CMB+Lensing+BAO 	&	$<$0.25	&	$<$3.84		\\
	 CMB+Lensing+BAO+Clustering 	&		$<$0.26		&	$<$3.80		\\
	 CMB+Lensing+BAO+Clusters 	&	$<$0.29	&$<$3.78		\\
	 CMB+Lensing+BAO+Shear  	&	$<$0.34		&	$<$3.79		 \\ 
	 CMB+Lensing+BAO+RSD  	&	$<$0.37		&	$<$3.75		 \\ \hline
	\end{tabular}
	\label{tab:active}
\end{table}
\begin{table}
	\centering 
	\caption{Same as Table~\ref{tab:active}, but for the $\Lambda$CDM+${\rm N}_{\rm eff}+\msterile$ model, showing tight constraints on the mass of sterile neutrinos.}
	\begin{tabular}{lccc}
		&	$\msterile$[eV] 	&	$\neff$ 	 	\\\hline
	 CMB+BAO	&	$<$0.28		&	$<$3.91	 	\\
	 CMB+Lensing+BAO 	&	$<$0.35	&	$<$3.84		\\
	 CMB+Lensing+BAO+Clustering 	&		$<$0.24		&	$<$3.87		\\
	 CMB+Lensing+BAO+Clusters 	&	$<$0.33		&$<$3.83		\\
	 CMB+Lensing+BAO+Shear  	&	$<0.51$	&	$<$3.82		 \\ 
	 CMB+Lensing+BAO+RSD  	&	$<0.59$	&	$<$3.70		 \\ \hline
	\end{tabular}
	\label{tab:sterile}
\end{table}

Tables~\ref{tab:active} and \ref{tab:sterile} summarise the constraints on neutrino masses in the active and sterile neutrino scenarios, respectively, {\it i.e.,} $\Lambda$CDM+${\rm N}_{\rm eff}+\msterile$ and  $\Lambda$CDM+${\rm N}_{\rm eff}+\msterile$ models, arising from a variety of data combinations. We see that multiple combinations yield similar constraints, and tend to small neutrino masses, {\it e.g.}, $\mnu$, $\msterile \lesssim 0.3$~eV at $95\%$ CL. Note that some of these constraints may be relaxed by adding freedom to the model, for example to the primordial power spectrum \cite{deputter_etal_2014}.
Interestingly, as also noted by Ref.~\cite{Beutler_etal_2014}, the Shear and RSD data prefer lower $\sigma_8$ and thus, larger neutrino mass. However, the Bayes factors presented in the second section of Table~\ref{tab:bayesfactors_tension} indicate a preference for the minimal $\Lambda$CDM model in all cases, even with the Shear and RSD data.  Note that Ref.~\cite{Beutler_etal_2014} marginalised over the lensing information which, as is well-known \citep{Planck_cosmo_2013}, leads to a preference for higher $\sigma_8$; conversely, our analysis combined the CMB temperature and lensing information. 

\begin{figure}
\hspace*{-2mm}\includegraphics[width=8.8cm]{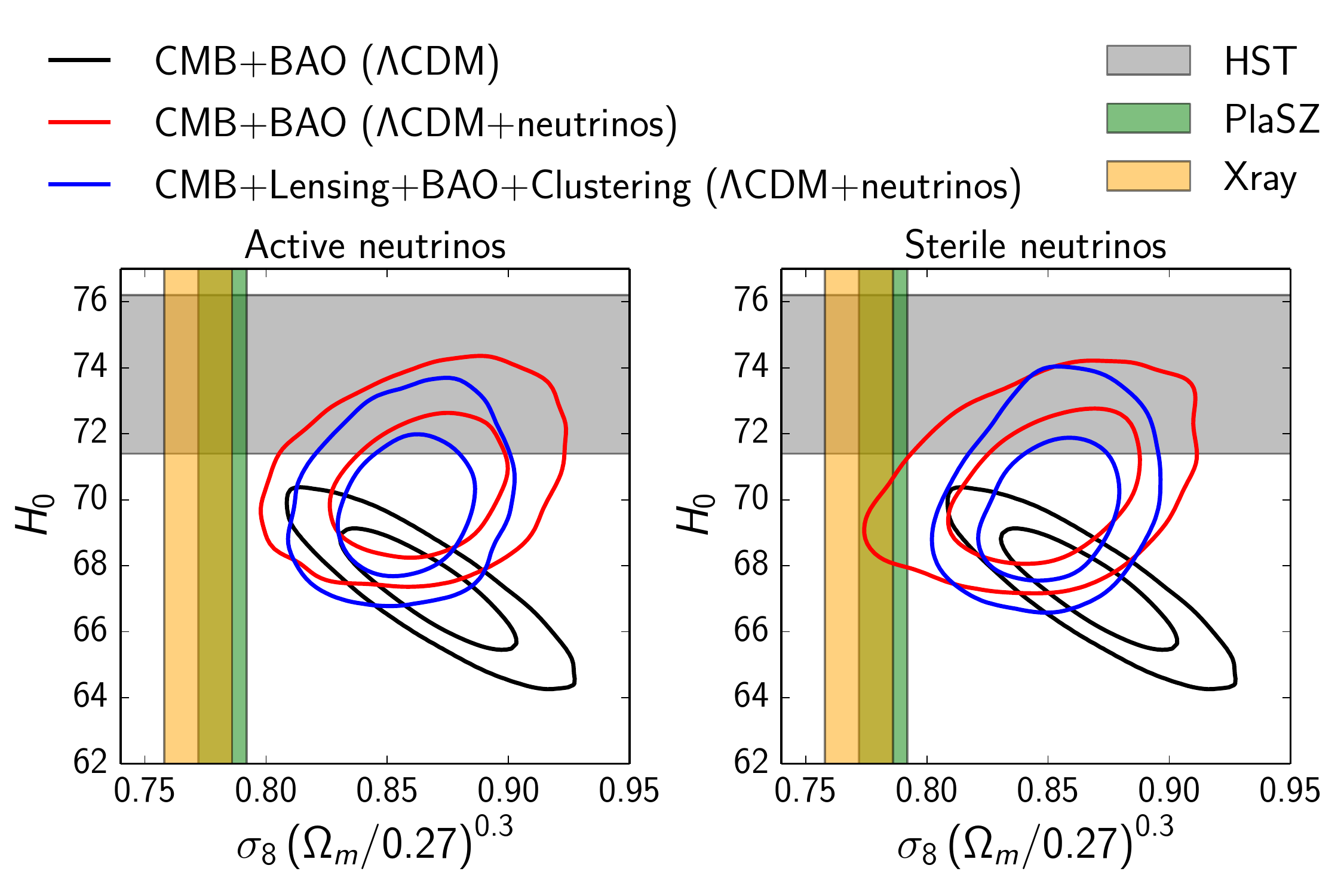}
\caption{Persistence of the tension as the minimal $\Lambda$CDM model is extended in the neutrino sector, {\it i.e.},  as ${\rm N}_{\rm eff}$ and massive active or sterile neutrinos are added. }
\label{fig:tensionboth}
\end{figure}

Fig.~\ref{fig:tensionboth} illustrates the persistence of the tension between the CMB+BAO, HST, PlaSZ and X-ray data, as one extends the minimal $\Lambda$CDM model in the neutrino sector. The tension with local measurements of $H_0$ is alleviated by ${\rm N}_{\rm eff}$ because of the degeneracy between these parameters \cite{Feeney:2013wp, Verde:2013cqa}, but the tension with PlaSZ and X-ray clusters persists despite the addition of both ${\rm N}_{\rm eff}$ and neutrino masses. The levels of tension are comparable in minimal and extended models when adding Lensing and Clustering data. We note that the PlaSZ and X-ray constraints were derived for the $\Lambda$CDM model, and it is unclear whether they can be used in the context of the extended models. In contrast, the datasets used in Tables~\ref{tab:active} and \ref{tab:sterile} all relied on uncompressed likelihoods or constraints shown to be usable within the extended models.

\begin{figure}
\hspace*{-6mm}\includegraphics[width=9cm]{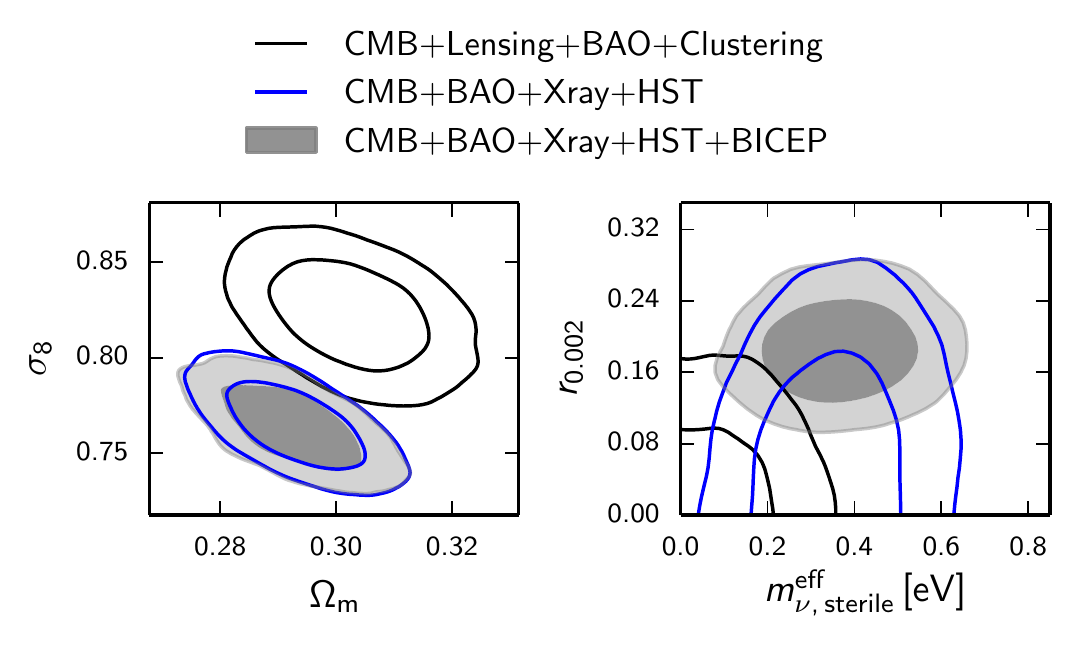}
\caption{Constraints on the $\Lambda$CDM$+\ r_{0.002}+{\rm N}_{\rm eff}+\msterile$ model, illustrating the persisting tension between X-ray clusters and CMB+BAO in the $\sigma_8$--$\Omega_m$ plane, despite an apparent reconciliation of the BICEP and \planck\ results on $r_{0.002}$. }
\label{fig:tensorcase}
\end{figure}

Finally, sterile neutrinos were claimed \cite{Zhang:2014dxk, Dvorkin:2014lea} to also resolve the tension in the \planck measurements of the tensor-to-scalar ratio ($r_{0.002}<0.11$ at $95\%$ CL) and the recent BICEP result, $r_{0.002} = 0.2^{+0.07}_{-0.05}$ \cite{bicep2_2014}. However, the tension in the $\sigma_8$--$\Omega_m$ plane detailed previously persists in the extended model  $\Lambda$CDM$+\ r_{0.002}+{\rm N}_{\rm eff}+\msterile$, as shown in Fig.~\ref{fig:tensorcase}. Hence, the relaxed constraints on $r_{0.002}$ from this data combination originates from a compromise between datasets in tension, not a new concordance. This is confirmed by the Bayes factors, presented in Table~\ref{tab:bayesfactors_tensortension}, showing that the extended model is not favoured over $\Lambda$CDM.

\textbf{Conclusions.} The need for extra parameters yielding a new cosmological concordance can only be convincing if the combined datasets are in tension in the minimal model, and in agreement in extended model. We show that massive sterile neutrinos do not bring about a new cosmic concordance, but rather highlight the tension between the CMB+BAO and SZ or X-ray clusters. A compilation of current LSS data which have been demonstrated to be robust to modelling uncertainties, when combined with {\it Planck}, tend to small masses $\mnu$, $\msterile \lesssim 0.3$~eV at $95\%$ CL in the context of the  $\Lambda$CDM model extended with $\neff$ and neutrino mass parameters. Similarly, as found in Refs.~\cite{Feeney:2013wp, Verde:2013cqa} the data cannot distinguish between $\neff \sim 3$ and $4$, and does not favour extra neutrinos  over the standard 3 families. These conclusions are corroborated by the Bayesian evidence:  the more complex models are not preferred, even when using datasets in tension. We conclude that current cosmological constraints do not provide evidence for large neutrino masses or extra neutrinos, even in the presence of the tension between \planck CMB and SZ and X-ray clusters. If this tension does not resolve after further investigation of systematic effects, new physics beyond massive neutrinos will be necessary to reconcile these datasets.

\begin{table}\def\arraystretch{1.1}
	\centering 
	\caption{ Evidence ratios $\ln [E_{\Lambda{\rm CDM}}/E_{\rm ext.}]$ between the minimal $\Lambda$CDM model and the  $\Lambda$CDM$+\ r_{0.002}+{\rm N}_{\rm eff}+\msterile$ model, showing that sterile neutrinos are not favoured by the data, even when adding the BICEP results.  }
	\begin{tabular}{lccc}
		  &	 Sterile	\\\hline
CMB+Lensing+BAO+Clustering  & $2.89  _{-0.19}   ^{+0.13} $ \\
CMB+BAO+Xray+HST  & $-0.70  _{-0.02}   ^{+0.07} $ \\
CMB+BAO+Xray+HST+BICEP  & $-0.66  _{-0.04}   ^{+0.05} $ \\
\hline
	\end{tabular}
	\label{tab:bayesfactors_tensortension}
\end{table}

\mbox{}

\acknowledgments

\paragraph{Acknowledgments.---}
We thank Ofer Lahav, Stephen Feeney, Nina Roth, Florian Beutler, August Evrard, Raphael Flauger, Marta Spinelli, and Filipe Abdalla, for useful discussions. BL is supported by the Perren Fund and the IMPACT Fund. HVP is supported by STFC and the European Research Council under the European Community's Seventh Framework Programme (FP7/2007- 2013) / ERC grant agreement no 306478-CosmicDawn. 
LV is supported by supported by the European Research Council under the European Community's Seventh Framework Programme FP7-IDEAS-Phys.LSS 240117 and Mineco grant FPA2011-29678- C02-02.  Based on observations obtained with {\it Planck} (http://www.esa.int/Planck), an ESA science mission with instruments and contributions directly funded by ESA Member States, NASA, and Canada.

\bibliography{tension}

\end{document}